\documentclass[10pt]{article}

\usepackage{graphicx}

\usepackage{amssymb}
\usepackage{amsxtra}
\title{\bf Contextual completeness and a classification scheme for theories}
\author{George Jaroszkiewicz \\
School of Mathematical Sciences, University of Nottingham, Nottingham, UK}

\begin{document}

\maketitle

\begin{abstract}
We discuss the role of propositions, truth, context and observers in scientific theories. We introduce the concept of generalized proposition and use it to define an algorithm for the classification of any scientific theory. The algorithm assigns a number 0, 1, 2 or 3 to a given theory, thereby classifying it as of metaphysical, mathematical, classical or quantum class. The objective is to provide an impartial method of assessing the scientific status of any theory.    

\end{abstract}

\section{Introduction}
\label{Section_1}

At a time when the empirical sciences are making ever greater advances in
quantum physics, it is a paradox that there should remain uncertainty and fruitless debate
regarding the interpretation of quantum mechanics (QM). This paradox has
been intensified by the speed and ease of modern communication. Everyone
thinks they are entitled to have an opinion that carries the same weight as any other: some say time is one of the
dimensions of the block universe, others say it is a process; some say
quantum wave functions collapse, others say they do not; some say there is
but one reality, others say there is an infinity of parallel worlds;
some say there are hidden variables, others say there are not; some say the
universe started as a quantum fluctuation, others ask \emph{of what?} and so
on. The great scientific principle of the Royal Society of London, \emph{%
	Nullius in verba} (take no-one's word for it) seems to have been abandoned
in many quarters in favour of unsupported opinion and pseudo-scientific
assertion. Hard won principles of science are routinely flouted in the media by experts who should know the difference between real science and science fiction.

It seems timely to assert that science is not a democracy but a
dictatorship: that of the laboratory. It is metaphysics and no business of
theorists to propagate as science unsubstantiated beliefs in undetectable parallel
worlds, or to advocate mathematical elegance as a scientific principle outweighing lack of empirical evidence, or
to assert that the universe originated in a quantum fluctuation, or to claim
that mathematical truth has the same status as physical observation %
\cite{TEGMARK-2014}. These are all baseless ideologies serving to breed a
complacent train of thought, no more useful in science than fairy stories.

\section{Truth values}
\label{Section_2}

The problem as we see it lies in the meaning of \emph{truth}. Whenever any
statement is made in whatever discipline, be it scientific, philosophical or
mathematical we should immediately ask several questions and insist on an
answer:

\begin{enumerate}
	\item {Has this statement been validated, i.e., assigned a truth value?}
	
	\item {Can it be validated? If so, by what actual means?}
	
	\item {Who has or will validate that statement, i.e., for whom or what is
		this statement intended to be meaningful?}
	
	\item {What would observers do with this statement if validated?}
\end{enumerate}

Outside of science, it has been fashionable for some time to abandon belief
in absolute truth and to regard truth as relative. However, relative truth
in science seems at face value to be inconsistent with empiricism: we want to believe that observations have a meaning beyond the narrow field of view of the observer. It is our
intention in this article to emphasize that the principle of relativity of
truth can indeed be placed on a proper scientific footing, fully in
accordance with empiricism. But it requires us to abandon belief in absolute scientific truth.

A by-product of our analysis is an effective
principle, a methodology, a weapon, that allows the expert and the
non-expert to relatively easily identify and classify assertions as either
metaphysical, mathematical, classical or quantum physical with a degree of
impartiality and certainty. We will propose such a principle, which we call 
\emph{contextual completeness}.

Contextual completeness allows us to examine any statement in a range of
subjects and assign a classification integer $k$ to that statement. If $k=0$, then we
can say we are dealing with a metaphysical statement; if $k=1$ we are
dealing with a mathematical or logical statement; if $k=2$ we are dealing
with a classical mechanical or classical relativistic statement; and if $k=3$
we are dealing with a quantum mechanical statement. For example, we find $%
k=0 $ for most statements in the Multiverse paradigm and the original
version of decoherence theory. Consistent histories scores $2$ at best.
Interestingly, frequentist probability theory scores 1 whilst Bayesian
probability scores $3$.

The principle of contextual completeness requires some explanation, or
context of its own. Therefore to understand our approach, we first clarify
our terms of reference, bearing in mind that what we are doing should be
thought of as \emph{metascience}, or \emph{how to do science}. It is not
precise in the preliminary form given here but should serve as a useful tool in
discussions about specific scientific theories. Hopefully it encourage a
more careful approach to the writing of science. Of course, our approach will
be the subject of criticism itself, but that is to be welcomed. Our test for contextual completeness can
be applied to comments as readily as to theories.

\section{Terms of reference}
\label{Section_3}

\subsection{Systems under observation}
\label{Subsection_3.1}

It is a powerful principle in science that the universe can be divided into
systems under observation (SUOs) and observers. There are two ways to think
about SUOs. The traditional and intuitive view is that SUOs ``exist'', have
physical properties and experiments can determine these properties. The
other view is the one expounded by Wheeler in his ``participatory principle" \cite{WHEELER-1979}:

\begin{quotation}
	``\emph{Stronger than the anthropic principle is what I might call the
		participatory principle. According to it we could not even imagine a
		universe that did not somewhere and for some stretch of time contain
		observers because the very building materials of the universe are these acts
		of observer-participancy. You wouldn't have the stuff out of which to build
		the universe otherwise. This participatory principle takes for its
		foundation the absolutely central point of the quantum: No elementary
		phenomenon is a phenomenon until it is an observed (or registered)
		phenomenon.}''\hfill J. A. Wheeler
\end{quotation}

From this perspective, SUOs cannot be discussed without the context of
observation: truth values have no meaning without observers to register
them. The Kochen-Specker theorem emphasises this point clearly %
\cite{KOCHEN-SPECKER-1967}.

\subsection{Observers}
\label{subsection_3.2}

A fundamental question in physics is: \emph{what is the essential difference
	between SUOs and observers?} We shall refer to this as \emph{the difference
	question}. The problem we have here is that according to the \emph{principle
	of the unity of physics}, SUOs and observers are described by the same laws
of physics. According to Feynman \cite{FEYNMAN-1982}:

\begin{quotation}
	``\emph{...we have an illusion that we can do any experiment that we want.
		We all, however, come from the same universe, have evolved with it, and
		don't really have any real freedom. For we obey certain laws and have come
		from a certain past}.'' \hfill R. P. Feynman
\end{quotation}

There is, according to this principle, no intrinsic difference between
observers and SUOs.

Whilst we entirely agree with this principle, it seems unsatisfactory
because it does not take into account the fact that observers and SUOs play
very different roles in physics. We may reconcile the difference question
and the unity of physics principle if as before we revise our notion of
truth and make it contextual. Given the difference question, we should ask
the crucially related question, \emph{who or what is interested in the
	difference?} \emph{To whom is the difference question addressed?}

The only answer that makes sense to us is: the difference question is
meaningful \emph{to the observer} and to no one or no thing else. The
difference between an observer and an SUO is \emph{contextual}. An observer
observes an SUO but that SUO does not observe the observer, \emph{relative
	to that observer}. It is perfectly possible however to imagine that an SUO $%
S_{1}$ being observed by an observer $O_{1}$ could act as an observer $O_{2}$
of $O_{1}$, in which case $O_{1}$ now plays the role of an SUO $S_{2}$, but
only relative to $O_{2}$.

When contextuality is respected, observers have characteristics that SUOs do
not have\footnote{%
	The essential question as to who notices this is answered contextually: the
	observer themselves.}. Observers have memories and an internal sense of
time, they build apparatus that can prepare states of SUOs, extract and
record information about those states, and eventually use it in some way
that is meaningful relative to those observers for purposes defined by those
observers. In context, SUOs do none of those things.

\subsection{Complete and partial observation}
\label{subsection_3.3}

Suppose two observers $O_{1}$ and $O_{2}$ observe an SUO $S$. There may well
be aspects of $S$ that $O_{1}$ observes that $O_{2}$ does not, and
vice-versa. In such a case each observer is observing $S$ partially. One of
the differences between classical mechanics (CM) and quantum mechanics (QM)
is that the principles of CM do not rule out \emph{complete observation},
that is, the notion that an observer can have available to them all possible
information about an SUO \emph{in principle}. QM on the other hand rules out
complete observation: there is no context in which an observer can observe
both the position and the momentum of a particle in one outcome. The famous
debate between Einstein, Podolsky and Rosen (EPR) on one side and Bohr on
the other boils down eventually to this difference between QM and CM %
\cite{EPR-1935,BOHR-1935}. Since QM does not allow complete, context-free
observation, there is no requirement for physicists to believe that it means
anything: we do not have to permit the EPR phrase ``element of reality'' into
the debate as legitimate.

\subsection{Primary observers}
\label{Subsection_3.4}

It is an empirical fact that observers do not have infinite lifetimes: they
come and go. Suppose we attempted to discuss the origin of some physical
observer $A$. This would be meaningful only from the perspective of some
earlier physical observer $B$ that had observed the creation or formation of 
$A$. But this would immediately raise the same question as to the origin of $%
B$, requiring some yet earlier observer $C$, and so in. If we allowed this
argument to continue indefinitely, we would be led to an infinite regress.
Such things are generally regarded as unsatisfactory.

To avoid such an infinite regress, we should stop somewhere, placing a veto
on questions about the origin of our chosen \emph{primary observer}. We stop
at some point convenient to us, accepting the existence of that observer for
granted and defining all contexts relative to them only and no further in.
This suggests that we may have to scale down our expectations of physics and
accept the possibility that we may never have a complete TOE (theory of
everything) or an understanding of the origin of the laws of quantum
mechanics.

This infinite regress problem was understood in antiquity. To avoid it,
Aristotle invoked the concept of an absolute primary observer known as a 
\emph{first mover} \cite{ARISTOTLE-PH}:

\begin{quotation}
	``\emph{Since motion must be everlasting and must never fail, there must be
		some everlasting first mover, one or more than one}.'' \hfill Aristotle
\end{quotation}

By definition, primary observers have a sense of time, memory and purpose,
for without any of these attributes they could not be regarded as observers.
At any moment of their time they hold data in their memories that they
interpret in terms of a hypothesized past relative to that moment. With
sufficient data, observers can even attempt to account for their own
origins. But that past is a map of the past and should not be confused with
it \cite{KORZYBSKI-1994}. Different primary observers might construct
different relative pasts and there is nothing in physics, apart from the
need for consistency should they exchange information, to prevent that. This
means that observer's past may be as uncertain for them as their future is.
It is metaphysics to assert that ``the past'' is unique and absolutely fixed.

It was the notable achievement of Hugh Everett III to enhance the discussion
about observers in QM. Specifically, he discussed the possibility of one
observer $O_{1}$ observing other observer $O_{0}$ who is performing
observations on some SUO $S_{0}$ \cite{EVERETT-1957}. All the rules of
standard QM are assumed to apply to the observations as made by $O_{0}$ on $%
S_{0}$. By the principle of the unity of physics, the same rules should
apply to the observations made by $O_{1}$ on the combined SUO $S_{1}$
consisting of $S_{0}$ and $O_{0}$.

How $S_{1}$ is related to $S_{0}$ and $O_{0}$ is a deep and interesting
question. Typically, $O_{0}$ describes $S_{1}$ in terms of a pure or mixed
quantum state. Also typically, $O_{1}$ describes $S_{1}$ as a pure or mixed
state involving tensor products of state vectors in a tensor product Hilbert
space.

Everett asserted that $O_{1}$ would/should regard $O_{0}$ and $S_{0}$
together as a single SUO, described by a wavefunction describing both as a
single SUO. Then, to avoid an infinite regress, Everett postulated that
there is a fundamental wavefunction, not just for our universe but for a
plethora of all possible universes. In addition, he asserted that only his
``Process $2$'' had intrinsic significance \cite{EVERETT-1957}:

\begin{quote}
	``The continuous, deterministic change of state of an isolated system with
	time according to a wave equation $\partial \psi /\partial t=A\psi $, where $%
	A$ is a linear operator.'' \hfill H.$\ $Everett$\ $III
\end{quote}

Everett gave no context of validity for this assertion, which
therefore represents a return to an absolutist perspective. This accounts
for the fierceness of the debate between those who believe that the Many
Worlds/Multiverse paradigm \cite{DEUTSCH-1997} is physics and those who see
it as metaphysics.

The concept of primary observer raises the question ``\emph{what then is the
	objective of physics?}'' If we are not allowed to ask questions about a
primary observer, surely that means that we have an incomplete understanding
of the universe. That is of course true. But our response is: \emph{does it
	matter?} Who said in the first place that quantum mechanics could explain
everything? Quantum mechanics only ever was a theory of \emph{observation},
not a \emph{description} of SUOs. The history of quantum mechanics is
littered with the debris of this debate. It was the realist Einstein who
attributed Planck's quanta to properties of the radiation field %
\cite{EINSTEIN-1905B}, whereas Planck himself only ever discussed the
responses of atomic oscillators in the walls of black body containers %
\cite{PLANCK-1900}. Although we think of Heisenberg's matrix mechanics %
\cite{HEISENBERG-1925} as formally equivalent to Schr\"{o}dinger's wave
mechanics \cite{SCHRODINGER-1926-01}, nothing could be further apart than
their respective views about what they had discovered.

At this point we emphasise that a primary observer is a contextual
construct: there is not a unique, `final' primary observer in the sense of
Aristotle above.

\subsection{Propositions}
\label{Subsection_3.5}

Arguably, the proper business of physics is to establish the truth status of
propositions or statements, denoted here generically by the symbol $P$.
Examples are $P_{1}\equiv $ `\emph{energy is always conserved'} and $%
P_{2}\equiv $ `\emph{the mass of the electron is} $9.10938291\times 10$
kilograms'. Propositions in the sense of the word employed here can be
elementary, in that they need not contain any conditionals such as `if', as
in the above examples $P_{1}$ and $P_{2}$. They can also be compound, such
as $P_{3}\equiv \ $`If it is Tuesday and it is eleven o'clock then if we
turn on this magnet, the electron beam will be deflected'.

\subsection{The validation function $\mathbb{V}$}
\label{Subsection_3.6}

We introduce the validation function $\mathbb{V}$, which maps all
propositions into the discrete set $\{0,1\}$. If we know $\mathbb{V}P=1$
then we know that $P$ is \emph{true}, whereas if $\mathbb{V}P=0$ then $P$ is 
\emph{false}.

\subsection{Context}
\label{Subsection_3.7}

The validation function $\mathbb{V}$ introduced above raises a
fundamental problem: we have used it without any information about the circumstances
that gave a truth value  $0$ or $1$. Also, there has been  no reference  to any observer for
whom the truth value is meaningful. Addressing this issue is the subject
matter of this article.

What is missing  is \emph{context}. In this article, a 
\emph{context} is all the information that gives meaning\footnote{%
	The question \emph{for whom is this meaningful?} will to be addressed
	presently.} to the validation process $\mathbb{V}$. Context should
include \emph{apparatus} when discussing physical propositions such as $%
P_{3}\equiv $ ``\emph{the momentum of this particle is }$\mathbf{p}$''.

Some reflection leads us to the conclusion that all propositions and laws of
physics are contextual. For instance, the proposition `\emph{energy is
	conserved}' is not an absolute truth: it does not hold in those situations
where energy can leak out of an otherwise closed system. It is metaphysics
to assert that energy is conserved for open systems: how could we know? The
truth status of the law of conservation of energy therefore is contextual.

Even the ``constants of physics'' are contextual. For instance, Planck's
constant lay undiscovered until the appropriate technology was available to
detect it. Even now, many observations in science can be explained in terms
of classical mechanics (CM), for which we set $\hslash =0$ for convenience.
We do not, as a rule, use QM to discuss the orbit of the Moon. Surprisingly,
even the proposition $P_{2}$ above is contextual: the effective mass of an
electron in vacuo is measurably different to the effective mass of an
electron in a crystal \cite{GAINUTDINOV-2012}. We point out that physicists
such as Dirac have explored the notion that physical `constants' depend on
time. If this is indeed the case, then these constants are contextual in a
deeper sense than simply contextual on our current technology and
mathematical modelling.

This raises an interesting question: can we create ``laws of physics'' by
devising contexts never hitherto seen in nature? For example, in Lagrangian
mechanics, we usually choose the Lagrangian $L$ to reflect the conditions in
the laboratory. For instance, if we have no external electromagnetic fields,
we might not have any electromagnetic potentials in $L$. Similarly, when the
search for the Higgs particle was under way, the presence of the Higgs field
in the Lagrange density was validated by the construction of very special
apparatus, which contributes to context. Another example is early universe
cosmology, where the effective `fundamental' Lagrangian changes radically as
the universe evolves from the Big Bang to the current epoch.

This line of argumentation does not lead to the conclusion that we can
create any law of physics that we like. Contrary to the Multiverse
conjecture that all things are possible in some universe, the universe we
live in does appear to have some rules. Empirical physics is the search for
those rules, for contexts that will validate certain physical propositions
and invalidate other. Given an arbitrary physical proposition, we cannot in
general always create contexts that will validate that proposition:
currently we cannot travel in time or exceed the speed of light.

\subsection{Varieties of proposition}
\label{Subsection_3.8}

Further consideration lead us to distinguish three classes of proposition: 
\emph{absolute}, \emph{contextual} and \emph{empirical}.

\subsubsection{Absolute propositions}
\label{Subsubsection_3.8.1}

An \emph{absolute proposition} $P$ is one given without any context. Such
propositions occur throughout metaphysics and philosophy and, unfortunately,
can be found in what is asserted to be science. The problem here is that,
for an absolute proposition to have a truth value, it has to have that truth
value relative to all conceivable contexts. But physicists cannot validate
any proposition relative to more than a finite number of contexts, simply
because every validation takes a finite time: the search for the Higgs boson
took many years. Therefore absolute propositions are meaningless in physics
because their truth status cannot be established.

\subsubsection{Contextual propositions}
\label{Subsubsection_3.8.2}

A contextual proposition consists of a proposition $P$ that is asserted
relative to a specific context $C$. We denote such a context $(P|C)$. A
contextual proposition can be true or false. If true we write $\mathbb{V}%
(P,C)=1$ whilst if false we write $\mathbb{V}(P,C)=0$.

In physics, a given proposition is generally true relative to more than a
single context. For example, let $P$ be the proposition $P\equiv $ `\emph{a
	free particle moves uniformly along a straight line}' (Galileo's law,
Newton's first law of motion), relative to the contexts $C_{1}\equiv \ $ 
`\emph{As observed in inertial frame} $F_{1}$' and $C_{2}\equiv \ $  `\emph{%
	As observed in inertial frame }$F_{2}$'. Then we have%
\begin{equation}
	\mathbb{V}(P_{,}C_{1})=\mathbb{V}(P,C_{2})=1 .
\end{equation}

Given a proposition $P$, we define its \emph{domain of validity} $\mathcal{C}%
_{P}$ to be the set of all contexts such that 
\begin{equation}
	C\in \mathcal{C}_{P}\Rightarrow \mathbb{V(}P,C\mathbb{)}=1.
\end{equation}

We define $\mathcal{C}^{\ast }$ to be the \emph{contextual universe}, the
class of all possible contexts\footnote{%
	We are not concerned here whether $\mathcal{C}^{\ast }$ is a set or a
	class.}. Then given $\mathcal{C}_{p}$, we define $\overline{\mathcal{C}}_{P}$
to be the complement of $\mathcal{C}_{P}$ relative to $\mathcal{C}^{\ast }$,
i.e. 
\begin{equation}
	C\in \overline{\mathcal{C}}_{P}\Rightarrow \mathbb{V}\mathbb{(}P,C%
	\mathbb{)}=0.
\end{equation}%
By definition, we have $\mathcal{C}^{\ast }=\mathcal{C}_{P}\cup \mathcal{%
	\bar{C}}_{P}$ and $\mathcal{C}_{P}\cap \mathcal{\bar{C}}_{p}=\phi .$

Suppose $C_{1}$, $C_{2}$, $\ldots $, $C_{n}$ are elements of the contextual
universe $\mathcal{C}^{\ast }$. Then we denote the set of simultaneous
contextual propositions

\begin{equation}
	(P|C_{1},C_{2},\ldots ,C_{n})\equiv (P|C_{1})\wedge (P|C_{2})\wedge \ldots
	\wedge (P|C_{n}),  \label{111}
\end{equation}%
where each of the terms $(P,C_{i}),i=1,2,\ldots ,n$, is a contextual
proposition and the symbol $\wedge $ is the logical $`$\emph{and'}. Then we
have the rule%
\begin{equation}
	\mathbb{V}(P,C_{1},C_{2},\ldots ,C_{n})=\displaystyle{%
		\prod\limits_{i=1}^{n}}\mathbb{V}(P,C_{i})
\end{equation}

The notation in (\ref{111}) does not mean that an experiment can be done
with two or more contexts simultaneously. Each factor on the right hand side
of (\ref{111}) refers to a separate contextual proposition. Contexts are
mutually exclusive: the factors that make up one context cannot be added to
a second context without creating a new context. This is one of the
fundamental rules of quantum mechanics that does not apply in classical
mechanics. Therefore, care has to be taken when contexts depend on
continuous parameters, for if a contextual parameter\footnote{%
	A contextual parameter is a parameter involved in the description of a
	context, such as the vertical angle of orientation of the main magnetic
	field in a Stern-Gerlach experiment.} is changed by what seems to be an
infinitesimal amount in some apparatus, the context and therefore the
experiment changes.

We defined an absolute proposition above as one that is independent of
context. Equivalently, its truth status is unaffected by choice of context,
so its domain of validity is the contextual universe. Therefore, if $P$ is
an absolute proposition, we may write%
\begin{equation}
	\mathbb{V}(P,C)=1\ \ \ \forall C\in \mathcal{C}^{\ast },
\end{equation}%
or equivalently,%
\begin{equation}
	P\equiv (P,\mathcal{C}^{\ast }).
\end{equation}

\subsubsection{Empirical propositions}
\label{Subsubsection_3.8.3}

An \emph{empirical proposition} is a contextual proposition that has been
validated under a sufficiently large number of contexts that it is regarded
as a law of physics, \emph{for all practical purposes} (FPP, %
\cite{PENROSE-1990}).

Any law of physics $L$ is necessarily an empirical proposition, because its
context set $\mathcal{C}_{L}\equiv \{C_{1},C_{2},\ldots ,C_{n}\}$ is always
necessarily a finite discrete set in physics, simply because every
validation experiment takes non-zero time. Assertions that involve
continuous parameters, such as `\emph{In each run of a Stern-Gerlach
	experiment, the electron spin value in any arbitrary orientation of the main
	magnet will always be plus or minus one half}', are theoretical abstractions
from a finite number of validated empirical propositions.

\section{Generalized propositions}
\label{Section_4}

We assert that absolute propositions are meaningless \emph{per se} in physics: we should always supply
a context for any physical proposition. However, that is still not enough.
For the truth value of a proposition $P$ to have a physical meaning, \emph{%
	two} conditions have to be satisfied, not one. These are $i)$ as stated
above, the proposition must have a context $C_{P}$ relative to which that
proposition might or might not be true, and $ii)$ the contextual proposition 
$(P,C_{P})$ must be \emph{contextually complete}. By this we mean that there
should be a statement identifying the \emph{primary observer} for whom that
contextual proposition's truth value is meaningful.

Our approach to classification of theories is to write every proposition in
a given theory as a \emph{generalized proposition. }If $P$ is a proposition, 
$C_{I}$ its context, and $O$ an observer, then its associated \emph{%
	generalized proposition} is of the form $(P,C_{I}|O,C_{E})$. Here $C_{E}$ is
the \emph{relative external context} that defines the observer $O$, relative
to whom the validity status of the proposition may be established and $C_{I}$
is the \emph{relative internal context} that includes details of the
apparatus used to validate the proposition \cite{J2010O}. If a generalized
proposition is true we write $\mathbb{V}(P,C_{I}|O,C_{E})=1$; if it is
false we write $\mathbb{V}(P,C_{I}|O,C_{E})=0$. Truth and falsity is
defined contextually by the observer and have no meaning outside of that
context.

\subsection{Contextual completeness}
\label{subsection_4.1}

The concept of context is a deep one and we have to deal with it on an
intuitive, heuristic basis. Relative internal context is generally easy to
deal with: if we were observing electron spin in a Stern-Gerlach experiment we
would need reasonably well-specified apparatus for instance. Relative
internal context may well involve abstract concepts such as mathematical
axioms, mathematical modelling of an SUO, and principles such as
wave-function superposition. Relative external context is generally more
familiar, as it will contain classical information that the observer has
about themselves, such as what sort of spacetime they are sitting in.

\subsection{Heisenberg cuts}
\label{subsection_4.2}

A \emph{Heisenberg cut} is a hypothetical line dividing the worlds of
quantum mechanics and classical mechanics. Heisenberg wrote \cite{HEISENBERG-1952}:

\begin{quotation}
	``\emph{The dividing line between the system to be observed and the
		measuring apparatus is immediately defined by the nature of the problem but
		it obviously signifies no discontinuity of the physical process. For this
		reason there must, within certain limits, exist complete freedom in choosing
		the position of the dividing line.}'' \hfill W.$\ $Heisenberg 
\end{quotation}

Although frequently discussed as if there an absolute divide between
classical and quantum worlds, Heisenberg cuts are contextual, separating
relative internal context and relative external context in any given quantum
mechanical generalized proposition. Given a contextually complete quantum
proposition $(P,C_{I}|O,C_{E})$, the corresponding Heisenberg cut is
represented by the vertical bar in our notation. Heisenberg cuts are
heuristic and impossible to define precisely, so occasionally there 
may be some uncertainty in or debate about our classification of a given
theory.

\subsection{Exophysics versus endophysics}
\label{Subsection_4.3}

An exophysical observer stands outside the SUO they are observing whilst an
endophysical observer is within a greater system. Some theories such as
Newtonian mechanics posit an absolute observer standing outside of space and
time, and is therefore an example of an exophysical observer. Exophysical
observers may be absolute, in that there is no relative external context
specified for them, or there may be such a context. For example, in
classical mechanics, we may associate a given inertial frame with an
exophysical observer at rest in that frame. Exophysical observers will be
denoted by the generic symbol $\Omega $ whilst endophysical observers are
denoted by $\omega $. Endophysical observers invariably have an associated
relative external context: otherwise, we would not know that they were endophysical.

In a given generalized proposition an absence of any context or reference to
an observer will be denoted by the empty set symbol $\emptyset $. For
example, an absolute proposition is denoted by $(P,\emptyset |\emptyset
,\emptyset )$.

\section{The classification of theories}
\label{section_5}

The various branches of knowledge such as metaphysics and philosophy,
mathematics and science can be classified in terms of degree of contextual
completeness. We can go so far as to assign a numerical value to this
classification as follows.

\subsection{The classification function}
\label{subsection_5.1}

We define the \emph{classification function} $\mathbb{K}$ that acts on
generalized propositions according to the rule%
\begin{equation}
	\mathbb{K}(P,C_{I}|O,C_{E})=\alpha +2\beta ,
\end{equation}%
where 
\begin{equation}
	\alpha =\left\{ 
	\begin{array}{c}
		0\ \text{if\ }C_{I}=\emptyset , \\ 
		1\ \text{if\ }C_{I}\neq \emptyset ,%
	\end{array}%
	\right. ,\ \ \ \beta =\left\{ 
	\begin{array}{c}
		0\ \text{if\ }C_{E}=\emptyset , \\ 
		1\ \text{if\ }C_{I}\neq \emptyset .%
	\end{array}%
	\right. .
\end{equation}%
The value $k\equiv \alpha +2\beta $ so obtained will be called the \emph{%
	classification} of the generalized proposition. We shall call generalized
propositions with classification \emph{0,1,2 }or\emph{\ 3} \emph{%
	metaphysical }or \emph{philosophical}, \emph{mathematical }or \emph{logical}%
, \emph{classical }or \emph{standard,} and \emph{quantum} or \emph{complete}
respectively.

A given theory may have generalized propositions of various classifications.
In such a case, the theory should be classified according to the
classification generally taken to characterize that theory. For example,
Hugh Everett's relative state theory has to be give a classification of $%
zero $ because Everett's core axiom ``\emph{Proposition 2}'' above makes no
reference to any observer: the `Many Worlds' wavefunction is asserted to
exist in an absolute sense. Indeed, B. de Witt, a leading advocate of Everett's work,
explicitly referred to the paradigm as metaphysical %
\cite{DeWITT-GRAHAM-1973}.

\section{Classifications}
\label{Section_6}

In this section we apply our classification scheme to a number of
disciplines and theories.

\subsection{Metaphysics}
\label{Subsection_6.1}

In metaphysics, typical generalized propositions are unverifiable
assertions of the form $(P,\emptyset |\emptyset ,\emptyset )$. Typical
metaphysical discussions involve the words \emph{existence, universals}, and
so on without giving any details as to how to validate such terms. Hence
typically $\alpha =\beta =0$, giving a classification $k=0$. Some branches
of metaphysics such as \emph{idealism }do attempt to relate to observers,
but not in any way that can be validated according to scientific principles.
Indeed, the classification of something as metaphysical is generally regarded as
synonymous with the impossibility of validation, and therefore,
unscientific. This is an important criticism of the Many worlds/Multiverse
paradigm discussed below.

We note that a generalized proposition that has a truth value of zero is not
a metaphysical one but a false mathematical, classical or contextually
complete proposition. Metaphysical generalized propositions have no truth
values.
 
\subsection{Philosophy}
\label{Subsection_6.2}

In a Philosophy not based on a specific moral code, typical
generalized propositions are contextually unsupported assertions relating to
some unspecified exophysical primary observer, which we write in the form $%
(P,\emptyset |\Omega ,\emptyset )$. Hence $\alpha =\beta =0$, giving
classification $k=0$.

In a Moral Philosophy based on a moral code $Code$, typical
generalized propositions are contextually supported assertions relating to
some unspecified primary observer, which we write in the form $%
(P,Code|\Omega ,\emptyset )$.

Such propositions can have contextual truth values: the function of $Code$
in this case is to provide a validation mechanism for a truth value of the
proposition $P$, ``truth'' here being
equated to being in accordance with the dictats of the code. Hence $\alpha
=1,\beta =0$, giving classification $k=1$.

\subsection{Logical Philosophy}
\label{Subsection_6.3}

In this discipline, a proposition $P$ is often a \emph{consequent}
of an \emph{antecedent }$A$, i.e., we encounter statements of the form $%
A\Rightarrow P$. If we interpret the antecedent $A$ as a sufficient
condition for the truth of $P$ then $A$ is part of the internal context. No
statement is made concerning any observer however, so generalized
propositions in this subject are usually of the form $(P,A|\Omega ,\emptyset)$. Hence $\alpha =1,\beta =0$, giving $k=1.$

\subsection{Mathematics and Formal Logic}
\label{Subsection_6.4}

In the traditional approach to mathematics, the truth value of a
proposition $P$ is determined by the axioms $Axioms$, i.e. we have%
\begin{equation}
	\mathbb{V}(Axioms\Rightarrow P)=1.
\end{equation}%
The axioms are a necessary part of the relative internal context. Standard
mathematics usually makes no explicit reference to any observer, although of
course mathematicians are aware that they are doing mathematics. At best we
have generalized propositions of the form $(P,Axioms|\Omega ,\emptyset ),$
so for standard mathematics, we find $k=1$. The same analysis applies to
most formal logic systems. Some branches of logic take some care in the
formal rules of the axioms, such as modal logic, but completely ignore the
question of who or what can take meaning from the results.

Plato's theory of forms explicitly demotes the role of any observer, be it
exophysical or endophysical, to an inessential role. Moreover, axioms are
not regarded as subject to any choice: they are just assumed to `be there'
as part of some higher universe of truth, whatever that means. This
automatically gives this interpretation of mathematics the metaphysical
classification of $k=0$. Validation when viewed as a decision process cannot
be applied to a proposition that is postulated to be true in the first
instance. This may explain why many working mathematicians are reluctant to
say whether they are Platonists or not: the mathematician who is seen to
merely stumble upon something already there seems not so creative as the
mathematician who invents a branch of mathematics.
 
There are several interesting exceptions to the classification $k=1$ for
mathematics:

\begin{enumerate}
	\item \textbf{Set theory:} Generally acknowledged as the Father of set theory, Cantor gave the following definition of a set \cite{CANTOR-1869}:
	
	`\emph{A set is a gathering together into a whole of definite, distinct
		objects of our perception or of our thought---which are called elements of
		the set}.' 
	
	Clearly, reference to \emph{perception} and \emph{our thought} relate to the mathematician describing the set, not to the actual set itself. The mathematician is at this point playing the role of an endophysical observer with a human mind, so we give Cantor's proposition a classification $k=2$. Clearly, if we rule out any reference to any
	thought processes, then Cantor's definition becomes contextually incomplete. Nowadays,
	mathematicians have generally discarded Cantor's approach and accepted that
	sets can be approached from the $k=0$ level, without any need for an
	observer. Apart from his definition of a set, Cantor's work on cardinality
	(the size of sets) rapidly reverted to $k=1$ mathematics, dealing with
	concepts of infinity that transcend any possible real observer to deal with
	or comprehend on an intuitive level. Semi-constructivists such as Kronecker
	found this profoundly objectionable.
	
	\item \textbf{Constructivist/intuitionist mathematics}
	
	The mathematicians who created both of these approaches to mathematics came
	to the conclusion that class $k=1$ mathematics begged certain questions. In
	the case of the constructivists, they believed that abstraction without the
	possibility of implementation is inadequate: they insisted that an observer
	has to have a model that implements the axioms of the mathematical system
	being studied. 
	
	The intuitionists believed that the truth of a mathematical statement is a subjective claim,
	corresponding to a mental construction of the mathematician. We identify
	such a mathematician as an observer. 
	
	Since both of these related approaches
	require some external context, such as a model or a sense of intuition, we
	classify them as class $3$ approaches to mathematics. 
	
	\item \textbf{The learning of mathematics:} our classification scheme plays
	a significant explicitly unstated role in the learning of mathematics, a
	role that is nevertheless well understood intuitively by teachers of the
	subject. Generally, students learning a new branch of mathematics find it
	hard to deal with a presentation at the $k=1$ level, as that is at a level
	of abstraction that the untrained human mind is not used to. The typical
	human tends to operate mentally at the solipsist level of thinking, for which
	propositions are dealt with at the $k=0$ level and the meaning of truth is debatable. 
	Mathematics is more effectively
	taught by examples, which are usually presented as a chain of steps taken by
	an observer. Students learn a new mathematical subject by copying steps from
	given examples, with the teacher assuming the role of an exophysical
	observer with experience. With sufficient practice, students then make a
	transition from a $k=2$ approach to mathematics to the more abstract $k=1$
	approach, i.e., a transition of the form%
	\begin{equation}
		(P_{1}\Rightarrow P_{2},\emptyset |\text{teacher,\ worked examples}%
		)\rightarrow (P_{2},\text{axioms}|\text{student},\emptyset ).
	\end{equation}
	
	\item \textbf{Non-euclidean geometry:} for two thousand years,
	mathematicians believed that propositions in geometry were of the form $%
	(P,E|\Omega ,\emptyset )$, where $E$ contains the context set of Euclid's
	five axioms and five postulates. Less than two hundred years ago, Gauss,
	Bolyai and Lobachewsky independently showed that generalized propositions in
	geometry could take the form $(P,E_{i}|\Omega ,C_{i})$, where $i=0$
	corresponds to the choice of Euclid's axioms and postulates, describing flat
	space, $i=1$ corresponds to the observer's choice of spherical geometry and $%
	i=3$ corresponds to hyperbolic geometry. The freedom to choose different
	geometries gives $k=3.$ Once the choice is made, we revert to $k=1.$
	
	\item \textbf{G\"{o}del's incompletness theorems: }Godel showed that for
	some axiomatic mathematical systems, an endophysical observer could add
	propositions that were apparently true but could not be validated from the
	axioms alone. There would be no way to disprove such propositions by finding
	counter examples, for instance. In our terms, we write%
	\begin{equation}
		\left\{ \mathbb{V}(P,\text{axioms}|\Omega ,\text{choice of }P)\overset{\text{%
				conjecture}}{=}1\right\} \nRightarrow \left\{ \mathbb{V}(P,\text{axioms}%
		|\Omega ,\emptyset )=1\right\}
	\end{equation}
	
	G\"{o}del's theorems implicitly involve an exophysical observer making
	choices, which is an attribute of a primary observer.
\end{enumerate}

\subsection{Computers, Church's $\lambda $ calculus and Turing machines}
\label{Subsection_6.5}

The advent of real computers illustrates the significance of our
classification system. An early approach to computation (in a formal sense
of the word) was that of the mathematician Alonzo Church, who developed an
approach to the study of computable functions involving a notation
developed by Church known as the $\lambda $-calculus \cite{PETZOLD-2008}.
Since Church's approach makes no reference to any observer, the $\lambda $%
-calculus has to be given the classification $k=1$, i.e., a purely formal,
mathematical one.

At about the same time, the theorist Alan Turing was developing what turned
out to be the same ideas, but he expressed them differently. Turing's
approach was to imagine an exophysical observer with a hypothetical machine,
now called a \emph{Turing machine}. Turing imagined that the observer
processed the same propositions that Church had considered, but via such a
machine. This elevates Turing's approach to a $k=3$ level. Moreover, the
Second World War stimulated the development of physical implementations of
Turin's concept, leading to the recent development of computers as we know
them now. If the Turing machine concept had not been developed, it is likely
that the world would have followed a radically different temporal development
in the previous century.

\subsection{Probability}
\label{Subsection_6.6}

Historically, there have been two conflicting approaches to
probability: the \emph{Frequentist} approach and the \emph{Bayesian}
approach. Our classification gives a different $k$ value to each.

The Frequentist approach takes the view that probability is an objective
attribute of an SUO and the measurement protocol. The Frequentist approach
does not deny the role of the latter: a fair dice that is dropped carefully
onto a table without any initial angular momentum and with its heads side
initially up should eventually come to rest with its heads side always up.
There is a role for the observer, who is assumed exophysical, but beyond
that, the observer is regarded as inessential. As far as probability is
concerned, the observer's role is to perform a number of runs or trials of a
basic procedure and count outcome values. In particular, the number of
trials or counts of frequency performed by the observer is assumed to be
sufficiently large so as to home in, in some way, on the perceived `absolute
probability' associated with the SUO and the sampling protocol. At that point the observer becomes irrelevant. Hence with Frequentism we
are dealing with generalized propositions of the form $(P,$ sampling
protocol$|\Omega ,\emptyset )$, giving a
classification of $k=1$. This approach to probability became more popular
with the advent of Kolmogoroff's axioms of probability, as these reinforce the mathematical side of probability.

On the other hand, the Bayesian approach to probability gives the observer
and their prior information a central role in the calculation of 
probabilities. This does not rule out an intrinsic (aleatoric) component to a
probability calculation. A typical generalized proposition in this approach
would take the form ($P$, aleatoric context $\arrowvert $ $\Omega $, observer's state of
knowledge), giving a classification $k = 3$.

The Bayesian approach to probability has on occasion been compared to the
way QM generates conditional probabilities. Attempts to re-interpret QM in terms of a
Bayesian approach to CM have not been successful because, whilst the
observer's ignorance is classical in both disciplines, relative internal
context in QM involves quantum mechanical rules. These cannot in general be duplicated 
by CM, as Bell inequalities have proven empirically \cite{BELL-1988}.

We note that contextual hidden variables theory attempts to link relative external context with relative internal context, but this has not led to any clear verdict apart from seeming to be a tortuous way to avoid QM.

\subsection{Classical Mechanics}
\label{Subsection_6.7}

CM emerged from the mists of antiquity and persisted into Renaissance times
in the form of Aristotelian mechanics. Generalized propositions in that
approach to motion would have been of the form $(P,\emptyset |\Omega
,\emptyset )$ relative to some unidentified absolute observer $\Omega ,$
giving a classification of $k=0$, i.e., metaphysics. Aristotelian
propositions about motion are generally contextually incomplete and conceptually flawed. Where they can be tested, such as the flight of projectiles, they are empirically
incorrect. 

Aristotelians appear to have been reluctant to contemplate any
empirical validation of their propositions on the grounds that their
objective was to understand the universe as it is. According to that
philosophy, experiments are intrinsically unnatural and any conclusions
drawn from them would not reflect the true universe.

Galileo is often regarded as the father of science on account of his
attitude towards empirical validation. He identified the significance of the
observer in the description of mechanics, and his ideas were encoded into
the first of Newton's laws of motion. Newton made considered statements
about absolute space and time and relative space and time in the Principia, %
\cite{NEWTON-1687} so generalized propositions in Galilean-Newtonian
mechanics are of the form $(P,\emptyset |\Omega ,$Absolute frame context).
This gives a classification $k=2$.

A further refinement was the realization that the absolute frame could not
be identified, so CM then became described in terms of exophysical observers
each at rest in their individual inertial frames. This leads to generalized
propositions of the form $(P,\emptyset ,|\Omega $, inertial frame context),
also giving $k=2$.

In general, varieties of CM have this classification, there being no real
discussion of the apparatus used to validate propositions. Indeed, the core
principles of CM are those of \emph{realism}, a belief in the objective
existence and reality of SUOs and their physical properties, all of which
are asserted to exist independently of any observers or processes of
observation.

The same realist agenda runs throughout Special Relativity and General
Relativity, so we give them the same CM classification $k=2$ when they are used to discuss physics. In case there is any doubt about this, we need only look at spacetime diagrams of black holes: these usually depict regions inside and outside the Schwarzschild radius using the same coordinate patch, as if physics could be discussed from the perspective of a superobserver standing outside spacetime. That is certainly a metaphysical perspective.
When they are described as branches of mathematics, SR and GR have the classification $k = 1$.

A proper quantum theory of gravity will only emerge when the role of observers and their apparatus is fully incorporated, and then we would expect a classification  $k = 3$, but the prospects at this time seem remote.

\subsection{Old Quantum Mechanics}
\label{Subsection_6.8}

We stated above that when in 1900 Planck postulated the
quantization of energy \cite{PLANCK-1900}, he was referring to the
oscillators in the walls of the black-body containers. Therefore, he was
discussing a crude model of apparatus, so generalized propositions are of
the form $(P$, detectors $|$ $\Omega $, inertial frame$)$, giving
a classification $k=3$.
The same analysis can be made of the Bohr-Sommerfeld atom. 

Despite its
ultimate failure to explain relative internal context, i.e., the
relationship between SUOs and apparatus, Old Quantum Mechanics (OQM) played
a pivotal role in the ultimate emergence of QM, as the older theory made a
break with the realistic classical principles that led to $k=2$ for CM. When Born proposed the probabalistic interpretation of Schr\"{o}dinger's wave function, he was already familiar with the application of probabilities to OQM.

\subsection{Einstein's photon theory}
\label{Subsection_6.9}

We also commented above that Einstein's association of quanta with the
electromagnetic field was a realist one, with no dependence on detectors.
Therefore, we classify his theory as $k=2$, i.e., classical. Einstein was
inherently a realist at heart despite his enormous contributions to the
development of quantum mechanics.

\subsection{Quantum mechanics}
\label{Subsection_6.10}

Bohr, Born, Heisenberg and Jordan cut their teeth on Old Quantum Mechanics and this
left its mark on their attitudes towards QM when it arrived in 1925. Bohr went on to develop the
Correspondence Principle, which links the quantum numbers associated with
relative internal context and the classical parameters of relative external
context. Therefore, we give the Correspondence Principle a classification $%
k=3$. 

Heisenberg developed Matrix mechanics on the basis that only what could be
observed was physically meaningful. Accordingly, we give a classification of 
$k=3$ to Matrix mechanics, Heisenberg's version of QM. On the other hand, Schr\"{o}dinger's
initial interpretation of his wave mechanics has to be given a
classification $k=2$ on account of his realist interpretation of the wave
function. The same classification gets assigned to hidden variables theories
such as Bohmian mechanics \cite{BOHM-1952}.

Schr\"{o}dinger's formalism was quickly interpreted in probabilistic terms
by Born, at which point it merits a classification of $k=3$. The
non-classical aspect of QM known as state reduction (wave function collapse)
involves changes in both relative internal and relative external context,
and not just the former and so the projection postulates associated with von Neumann
and L\"{u}ders get classified as $k = 3$.

At this point our classification scheme may be inadequate to do full justice to certain quantum theories. We refer to those applications of quantum theory that attempt to model both SUOS and apparatus specifically. We have in mind particularly the notable paper by Mott on the detection of alpha particles in a cloud chamber \cite{MOTT-1929} and Unruh's paper on accelerating detectors in Minkowski spacetime \cite{UNRUH-1976}, as well as the  work of Schwinger, Glauber and others on localized detection models.
We would assign a classification $K = 3^{\ast}$ to those theories.

\subsection{Thermodynamics and statistical mechanics}
\label{Subsection_6.15}

Classical theoretical thermodynamics views entropy as a property of an SUO, which would on that basis give the subject a classification $k =0$. However, the subject also discusses irreversible changes in entropy, which involves the physical time associated with the external observer. Therefore, we give the subject a classification $k = 2$.

Classical statistical mechanics introduces probability into the observer's relative external context, but does not require a discussion of apparatus in the truth values of its generalized propositions. Therefore we assign it a classification of $K =2$ also.

Quantum statistical mechanics involves both internal relative context and relative external context for its truth values, giving a classification $k = 3$.

\subsection{Quantum field theory}
\label{Subsection_6.11}

As an application of quantum principles to SUOs with continuously
many degrees of freedom, relativistic quantum field theories based on
Minkowski spacetime have a classification of $k=3$. Several perspectives
support this conclusion, although apparatus is rarely explicitly modelled.
One important feature that helped establish the empirical validity of
quantum electrodynamics was the success of the renormalization programme.
This depends on a heuristic bridge between the divergences associated with
the relative internal context (involving metaphysical concepts such as bare
mass and bare fields) and the relative external context associated with
physical values of particle mass and other observables. Another feature is
the explicit reference to the \emph{in} and \emph{out} states of the LSZ
scattering formalism \cite{LSZ-1955}, interpreted here as part of the
external relative context. Yet another feature supporting the classification 
$k=3$ is the embedding of Lorentz and other symmetries into the unitary
evolution formalism associated with the relative internal context and the
frame transformation properties associated with the relative external
context.

Attempts to extend the formalism to more general background spacetimes runs
into technical problems on all fronts: the renormalization program is poorly
defined and there is no natural concept of \emph{in} and \emph{out} states.
Even the quantized particle concept has issues in curved spacetimes \cite{COLOSI+ROVELLI-2009}.

\subsection{Quantum gravity and quantum cosmology}
\label{Subsection_6.12}

The ease and frequency with which quantum gravity and quantum
cosmology are discussed in the imaginary time formulation suggests that
these topics are not manifestly based on empiricism. Indeed, quantum gravity
has had a long history during which the identification of the correct
observables to quantize had no consensus and there is currently no way that
any of its concepts can be validated empirically. Quantum cosmology has gone
even further into the realms of metaphysics: the Wheeler-de Witt equation
suggests that the quantum state of the universe does not evolve in time
(assuming such a concept has any more than a metaphysical significance). 

We assign a classification $k = 0$ to both of these subjects because whilst they are mathematical in general form, they purport to be theories of physics. 

\subsection{M-theory}
\label{Subsection_6.13}
 
This is manifestly metaphysics, with barely comprehensible propositions asserted with  no reference to any relative internal or external context or observer whatsoever. Even the origin of the latter ``M'' is uncertain, so we would propose it to be pinned down as \emph{Metaphysics}, in line with its classification.

\subsection{Experimental reports}
\label{Subsection_6.14}

Experiments of all types represent the ultimate in proper scientific
activity. A complete report of an experiment will be contextually
complete, documenting all relevant validation contexts, such as a statement of the
problem, a description of the apparatus and the experimental protocol, an
identification of the individuals carrying out the experiment and tables of
all results. Important additional information that should always be given is
the time and place at which the experiment was done, because this can have a
significant effect on outcome, as the following examples illustrate:

\begin{enumerate}
	\item In a rotating frame of reference such as the earth, any application of
	Newtonian mechanics should take into account frame-induced forces such as
	`centrifugal' force;
	
	\item In the Large Hadron Collider, the position of the moon affects the
	calibration of the detectors and this has to be adjusted for %
	\cite{GAGNON-2012};
	
	\item The universe is expanding, according to consensus. Therefore, there
	would have been an epoch in the early universe when atoms did not exist.
	Discussions of early universe big bang temperatures, pressures and particle
	densities are therefore necessarily contextually incomplete on that account.
	Therefore, early universe cosmology should be carefully discussed so as to
	emphasize the fact that all its propositions are contextual on current
	observations and theoretical retrodiction based on them. This is reasonable
	as far as classical mechanics is concerned but in the case of quantum
	processes goes against the Wheeler's dictum, the last sentence in the quote from Wheeler given
	above. This warns us not to use counterfactuality when discussing quantum
	mechanics. This will be particularly relevant when discussing hypothetical
	objects such as ``quantum black holes'' in the early universe.
\end{enumerate}

\section{Concluding remarks}
\label{Section_7}

We emphasize that relative internal context does not refer to a description of a proposition, but to the mechanism, apparatus, or set of axioms relative to which that proposition can be validated. For example, if we assert that `the pressure in this box is one atmosphere'', the box would not be listed as part of the relative internal context. To give that context we would have to explain how we could establish the truth of that proposition, i.e., we would have to define the pressure meter involved and details of where it was placed in the box, and so on.

Our classification of theories is at best a guide to the significance of
various theories. Undoubtedly it could be improved on. However, it may have
a value in identifying research initiatives that may be 
``not even wrong''. For example, GR is a perfectly
respectable and reputable classical theory with $k=2$. Generalized
propositions in GR are of the form (P,$\emptyset |\Omega $, metric). Attempts
to quantize GR would require generalized propositions of the form $(P,$%
observables$|\omega $, localized laboratory context) with $k=3$. It has
proven difficult to see how to make the transition from classical to quantum
in this context in a convincing or meaningful way. Although the technical
problems seem insurmountable, the conceptual issues in quantum gravity are
the real problem. A classification approach such as the one we have outlined
may be a useful tool to have.

Finally, we give a useful rule. In the application of the contextual completeness test of any proposition, the most useful questions to ask of the author of that proposition are: \emph{how do you know this to be true}? and \emph{who are you talking to}? It is interesting to discover how many propositions are made in nominally scientific articles where no answers are forthcoming.

\bibliographystyle{plain}      

\end{document}